\documentclass[11pt]{article}
\usepackage{epsfig,amsmath,amssymb,mathtools,amsthm}
\usepackage{euscript}
\usepackage[authoryear,round]{natbib}

\usepackage[usenames,dvipsnames]{color}
\usepackage{pdflscape}
\usepackage{multirow}
\usepackage{booktabs}
\usepackage{threeparttable}
\usepackage{adjustbox}
\usepackage{tablefootnote}
\usepackage{graphicx}
\usepackage{subcaption}

\usepackage[table]{xcolor}
\usepackage{colortbl}

\makeatletter

\newcommand{\be}{\begin{eqnarray}}
\newcommand{\ee}{\end{eqnarray}}
\newcommand{\ben}{\begin{eqnarray*}}
\newcommand{\een}{\end{eqnarray*}}

\newtheorem{theorem}{Theorem}

\begin{document}

\title{\bf Modeling large dimensional matrix time series with partially known and latent factors\footnote{
 This research is supported by National Social Science Foundation
 No.23BTJ057. } 
}

\author{Yongchang Hui \\ {\small Xian Jiaotong University, Xian, China
} \\
Yuteng Zhang \\
{\small Xian Jiaotong University, Xian, China} \\
Siting Huang \\
{\small Xian Jiaotong University, Xian, China
}
}
\date{November, 2024}
\maketitle

\begin{abstract}
This article considers to model large-dimensional matrix time series by introducing a regression term to the matrix factor model. This is an extension of classic matrix factor model to incorporate the information of known factors or useful covariates.
We establish the convergence rates of coefficient matrix, loading matrices and the signal part. The theoretical results coincide with the rates in \citet{wang2019factor}. We conduct numerical studies to verify the performance of our estimation procedure in finite samples. Finally, we demonstrate the superiority of our proposed model using the daily returns of stocks data.
\end{abstract}

\vspace{3em}

\newpage

\section{Introduction}
\noindent{}With advancements in data collection technologies, the utilization of large-scale matrix time series data has surged across various fields, including finance, economics, healthcare, and environmental science. For instance, scholars have extensively examined the monthly returns of 100 portfolios, organized by ten levels of market capital size and book-to-equity ratio (e.g. \citet{wang2019factor}; \citet{chen2020constrained}; \citet{yu2022projected}; \citet{he2022matrix}). Moreover, data sources such as the international trade flow data (\citet{chen_chen_2022}), the multinational macroeconomic indices data (\cite{chen2023statistical}), the credit default swap prices data (\citet{yu2024matrix}) have been extensively studied. 
Consequently, the modeling and analysis of such complex and voluminous data sets pose significant challenges. 

To address these issues, \citet{wang2019factor} proposed the matrix factor model, which maintains the inherent correlations between rows and columns of the data and achieves dimension reduction in two directions. \citet{chen2023statistical} considered a different setting of matrix factor model by allowing the weak correlation across observations in the error process. They proposed an $\alpha$-PCA method, which is applicable to independent identically distributed (iid) and weakly auto-correlated matrix data. Follow this line of research, \citet{yu2022projected} proposed a projection estimation method that makes full use of the joint low rank structure across both rows and columns of matrix factor model. To handle heavy-tailed data, \citet{he2024matrix} pursued robustness by minimizing the Huber loss function to obtain estimators for the matrix factor model. Other developments and extensions include \citet{chen2020constrained}, who  expanded the model to its constrained version, and \citet{liu2022identification}, who extended it to the threshold matrix factor model. Additionally, \citet{gao2023two} proposed a two-way linear transformation factor model, and 
\citet{chang2023modelling} based on tensor canonical polyadic (CP)-decomposition to model matrix time series.

Despite the rapid development of matrix factor models in recent years, current research overlooks an important fact: the existence of known factors, particularly in the field of finance.  These known factors  not only provide valuable information in modeling the data, but also enhance the interpretability of the models.  The importance of including known factors into financial models has been extensively demonstrated in the literature. For instance, the  Capital Asset Pricing Model (CAPM) studied by \citet{sharpe1964capital} provides a extensive framework for understanding the relationship between risk and return. Building on this, \citet{fama1993common} introduced the three-factor model, which more effectively explains the cross-section of average stock returns. Later,  \citet{fama2015five} expanded this framework to a  five-factor model, further enhancing its explanatory power. Additional significant works include those by \citet{carhart1997persistence}, \citet{sahay2005multi}, \citet{psychogios2012lean}, \citet{ross2013arbitrage}, \citet{tian2015automated}, etc. All these studies demonstrate  the significance of incorporating known factors into financial models to improve both their explanatory power and practical applicability. 
Inspired by prior literature, we propose the following model for observations $\mathbf{Y}_t \in \mathbb{R}^{p\times q},\ 1\leq t\leq T$:
\begin{equation} \label{E1}
    \mathbf{Y}_t = \mathbf{A} \mathbf{X}_t + \mathbf{R} \mathbf{F}_t \mathbf{C}^{\prime} +\mathbf{E}_t, 
 \end{equation}
 where $\mathbf{X}_t$ is an observable $m \times q$ time series, 
 $\mathbf{F}_t $ is the $k \times r$ unobserved time series of common fundamental factors, $\mathbf{E}_t$
 is a $p \times q$ error matrix, $\mathbf{A}$ is a $p\times m$ unknown regression coefficient matrix, $\mathbf{R}$ and 
 $\mathbf{C}$ are unknown front loading matrix and back loading matrix, respectively. We assume that vec($\mathbf{E}_t$) $\sim$ WN($0, \Sigma_{\varepsilon})$ 
and vec($\mathbf{E}_t$) is uncorrelated with (vec($\mathbf{X}_t$), vec($\mathbf{F}_t$)). The form of the first term in model \eqref{E1} follows the idea of \citet{viroli2012matrix} and \citet{mao2019matrix}. Each column of $\mathbf{X}_t$ can be interpreted as the known factors corresponding to the respective column of $\mathbf{Y}_t$ as shown in Section \ref{Section2}. When there is no information of $\{\mathbf{X}_t\}$, our model reduces to the classic matrix factor model proposed by \citet{wang2019factor}. Perhaps a more natural way to incorporate both known and unknown factors into a matrix factor model is through the following formulation:
    $\mathbf{Y}_t = \mathbf{A} \mathbf{X}_t \mathbf{B}^{\prime} + \mathbf{R} \mathbf{F}_t \mathbf{C}^{\prime} +\mathbf{E}_t,$ 
where $\mathbf{X}_t$ is an observable $m \times n$ time series. However, empirical studies  on matrix-variate time series are still in its early stages (\citet{chen2023statistical}; \citet{he2023one}). Currently, there are no widely recognized matrix-variate known factors in any application scenarios. For instance, \citet{wang2019factor} extracted a $2 \times 2$ matrix factor to analyze stock return data; however, they acknowledged that further analysis is needed to validate the effectiveness of this matrix factor. 

Related to this work is a substantial body of literature on semi-parametric time series models, as each column of $\mathbf{X}_t$ can also be regarded as a vector of covariates associated with the corresponding column of $\mathbf{Y}_t$. \citet{engle1986semiparametric} proposed the partially linear semi-parametric model to analyze the relationship between temperature and electricity sales. \citet{gao2006estimation} developed an estimation procedure for the semi-parametric additive time series model and \citet{xia1999extended} analyzed the semi-parametric single-index model. Other influential work included those by \citet{robinson1988root}, \citet{fan1999root}, \citet{gao2004semiparametric}, \citet{davidson2002establishing} and \citet{li2015flexible}. Also \citet{gao2007nonlinear} provided a comprehensive review of these semi-parametric time series models.   Early attempts involved incorporating covariates information into traditional vector factor model, resulting in the development of semi-parametric factor models (\citet{connor2007semiparametric}; \citet{connor2012efficient}; \citet{fan2016projected}), and introducing regression terms with known factors to the factor model (\citet{chang2015high}).

For the proposed model, this paper presents a two-step estimation procedure. In the first step, we apply the least square method to obtain the estimation of coefficient matrix $\mathbf{A}$. This allows us to transform model \eqref{E1} into a traditional matrix factor model. Consequently, the estimation procedure introduced by \citet{wang2019factor} can be used in the second step to estimate the factor loadings and determine the dimensions of latent factors. On theoretical aspects, we establish the convergence rates for the coefficient matrix and factor loading matrices.  Notably, the convergence rates of factor loading matrices are the same as in \citet{wang2019factor}.


The rest of the paper is organized as follow. We introduce the proposed model in Section 2. In Section 3, we present our estimation procedure. The theoretical properties of the proposed estimators are investigated in Section 4. Section 5 studies the finite sample performance through simulation. A real data example is analyzed in Section 6. Section 7 concludes.


\section{Model} \label{Section2}
\noindent{}To begin, we first discuss the identifiable issue existing in model \eqref{E1}. That is, for any $k \times k$ 
and $ r \times r$ invertible matrices $\mathbf{U}_1$ and $\mathbf{U}_2$, model \eqref{E1} is exactly the same by substituting the triplets $(\mathbf{R}, \mathbf{F}_t, \mathbf{C})$ with $(\mathbf{R} \mathbf{U}_1, \mathbf{U}_1^{-1} 
\mathbf{F}_t \mathbf{U}_2^{-1} , \mathbf{C} \mathbf{U}_2^{\prime})$. However, as shown in \citet{lam2011estimation} and \citet{wang2019factor}, the column spaces of the factor loadings $\mathbf{R}$ 
and $\mathbf{C}$, denoted by $\mathcal{M} (\mathbf{R})$ and $\mathcal{M} (\mathbf{C})$ respectively, are uniquely determined. Based on QR decomposition, $\mathbf{R}$ and $\mathbf{C}$ can be represented as:
$$
 \mathbf{R} = \mathbf{Q}_1 \mathbf{K}_1,\quad \text{and}\quad \mathbf{C} = \mathbf{Q}_2 \mathbf{K}_2, 
$$
where $\mathbf{Q}_i$ is semi-orthogonal matrix and $\mathbf{K}_i$ is non-singular matrix, for $i=1,2$. So, $\mathcal{M}(\mathbf{R}) = \mathcal{M }(\mathbf{Q}_1)$ and $\mathcal{M}(\mathbf{C}) = \mathcal{M}(\mathbf{Q}_2)$. Then the model \eqref{E1} can be rewritten as 
\begin{equation}
    \mathbf{Y}_t = \mathbf{A} \mathbf{X}_t + \mathbf{Q}_1 \mathbf{K}_1 \mathbf{F}_t \mathbf{K}_2^{\prime} \mathbf{Q}_2^{\prime} + \mathbf{E}_t = \mathbf{A} \mathbf{X}_t + \mathbf{Q}_1 \mathbf{Z}_t \mathbf{Q}_2^{\prime} + \mathbf{E}_t,
\end{equation}
where $\mathbf{Z}_t = \mathbf{K}_1 \mathbf{F}_t \mathbf{K}_2^{\prime}.$

To facilitate comprehension of the model \eqref{E1}, we present an interpretation that can be attained through the following three steps.


\vspace{0.2in}
\noindent{\bf Interpretation:}

\vspace{0.1in}
\noindent{\bf Step 1:} For each fixed column $j=1,2,\ldots,q$, using data $\{\mathbf{y}_{t,\cdot j} =(Y_{t,1j},\ldots,Y_{t,pj})^{\prime} ,\ t=1,2,\ldots,T\}$ to fit a typical factor model. Assuming there are $m$ dimensional known factors $\{\mathbf{x}_{t,\cdot j}=(X_{t,1j},\ldots,X_{t,mj})^{\prime},\ t=1,2,\ldots,T\}$ with a $p\times m$ dimensional loading matrix $\mathbf{A}^{(j)}$ and $k_1$ dimensional unknown factors $\{\mathbf{g}_{t,\cdot j}=(G_{t,1j},\ldots,G_{t,k_1 j})^{\prime},\ t=1,2,\ldots,T\}$ with a $p\times k_1$ dimensional loading matrix $\mathbf{R}^{(j)}$. That is,
\begin{equation*}
     \begin{pmatrix}
       Y_{t,1j} \\
       \vdots \\
       Y_{t, p j}
   \end{pmatrix} = \mathbf{A}^{(j)}  \begin{pmatrix}
       X_{t,1j} \\
       \vdots \\
       X_{t, m j}
   \end{pmatrix} + \mathbf{R}^{(j)}
   \begin{pmatrix}
       G_{t,1j} \\
       \vdots \\
       G_{t, k_{1} j}
   \end{pmatrix}
   + \begin{pmatrix}
       H_{t,1j} \\
       \vdots \\
       H_{t,p j}
   \end{pmatrix} ,\quad  t=1,2,\ldots,T.
\end{equation*}
Let $\mathbf{X}_t$, $\mathbf{G}_t$, and $\mathbf{H}_t$ represent matrices with the $j$th column being $\mathbf{x}_{t,\cdot j}$, $\mathbf{g}_{t,\cdot j}$, and $\{\mathbf{h}_{t,\cdot j}=(H_{t,1j},\ldots,H_{t,p j})^{\prime}\}$, respectively.

\vspace{0.1in}
\noindent{\bf Step 2:} Suppose each row of $\mathbf{G}_t$ also assumes the factor structure,  with a $k_2$ dimensional factors $\{\mathbf{f}_{t,i \cdot }=(F_{t,i1},\ldots,F_{t,i k_2 }),\ t=1,2,\ldots,T\}$ and a $q \times k_2$ loading matrix $\mathbf{C}^{(i)}$. That is,
\begin{equation*}
    (G_{t,i 1},\ldots,G_{t, i q})  = (F_{t,i 1},\ldots,F_{t, i k_2}) \boldsymbol{C}^{(i)^{\prime}} + (H^*_{t, i1},\ldots,H^*_{t, i q}), \quad i=1,\ldots,k_1, t=1,\ldots,T.
\end{equation*}
Let $\mathbf{F}_t$ and $\mathbf{H}_t^{*}$ represent matrices with the $i$th row being $\mathbf{f}_{t,i \cdot}$ and $\{\mathbf{h}_{t,i \cdot }^* =(H_{t,i1},\ldots,H_{t,i q})\}$, respectively.

\vspace{0.1in}
\noindent{\bf Step 3:} If we assume $\mathbf{A}^{(1)} = \cdots= \mathbf{A}^{(q)}= \mathbf{A}$, $\mathbf{R}^{(1)} = \cdots= \mathbf{R}^{(q)}= \mathbf{R}$ and $\mathbf{C}^{(1)} = \cdots= \mathbf{C}^{(k_1)}= \mathbf{C}$, we have  $\mathbf{Y}_t= \mathbf{A} \mathbf{X}_t + \mathbf{R} \mathbf{G}_t + \mathbf{H}_t,$ and $\mathbf{G}_t= \mathbf{F}_t \mathbf{C}^{\prime} +  \mathbf{H}_t^{*}$. Hence
\begin{equation*}
    \mathbf{Y}_t= \mathbf{A} \mathbf{X}_t + \mathbf{R} \mathbf{F}_t \mathbf{C}^{\prime} + \mathbf{R} \mathbf{H}_t^* + \mathbf{H}_t = \mathbf{A} \mathbf{X}_t + \mathbf{R} \mathbf{F}_t \mathbf{C}^{\prime} +\mathbf{E}_t,
\end{equation*}
where $\mathbf{E}_t = \mathbf{R} \mathbf{H}_t^* + \mathbf{H}_t$. It is corresponding to model \eqref{E1}.

Based on the above interpretation, it can be inferred that each column of the matrix time series $\{ \mathbf{X}_t\}$ can be regarded as the known factors for each column of the matrix time series $\{ \mathbf{Y}_t\}$.

\noindent
{\bf Remark 1:\ }
    When the response is a $p \times 1 $ vector, the model \eqref{E1} becomes 
\begin{equation}
    \mathbf{y}_t = \mathbf{A} \boldsymbol{x}_t + \mathbf{R} \mathbf{f}_t + \mathbf{e}_t,\quad t=1,...,T,
 \end{equation}  
where $\mathbf{x}_t$ is an observable $m\times 1$ time series, $\mathbf{f}_t$ is a $k \times 1$ latent factor process. It corresponds to the model studied by \citet{chang2015high}. Thus, the model \eqref{E1} can be viewed as a generalization of the model proposed in \citet{chang2015high}.
\medskip

\section{Estimation} \label{S3}
\noindent{}Our objective is to estimate the unknown coefficient matrix $\mathbf{A}$, along with the column spaces $\mathcal{M}(\mathbf{Q}_1)$ and $\mathcal{M}(\mathbf{Q}_2)$, and 
the latent dimensions $(k,r)$. We begin with the estimation of matrix $\mathbf{A}$.
Denote
\begin{equation}
    \mathbf{W}_t = \mathbf{R} \mathbf{F}_t \mathbf{C}^{\prime} + \mathbf{E}_t = \mathbf{Q}_1 \mathbf{Z}_t \mathbf{Q}_2 + \mathbf{E}_t.
\end{equation}
Suppose that the factor process $\mathbf{F}_t$ and $\mathbf{X}_t$ are uncorrelated, i.e., 
$\text{Cov}(\text{vec}(\mathbf{X}_t), \text{vec}(\mathbf{F}_t)) = \mathbf{0}$, then we have
\begin{equation*}
    \text{Cov}(\text{vec}(\mathbf{X}_t), \text{vec}(\mathbf{W}_t)) = \text{Cov}(\text{vec}(\mathbf{X}_t), (\mathbf{C} \otimes \mathbf{R})\text{vec}(\mathbf{F}_t)) + \text{Cov}(\text{vec}(\mathbf{X}_t), \text{vec}(\mathbf{E}_t)) = \mathbf{0} .
\end{equation*}
Therefore, the estimation of $\mathbf{A}$ can be treated as a standard least square problem. The least squares estimator of $\mathbf{A}$ is given by 
\begin{equation*}
    \widehat{\mathbf{A}}=(\frac{1}{T} \sum_{t=1}^{T} \mathbf{Y}_t \mathbf{X}_t^{\mathrm{T}})(\frac{1}{T} \sum_{t=1}^{T} \mathbf{X}_t \mathbf{X}_t^{\mathrm{T}})^{-1} .
\end{equation*}
Consequently, we obtain a natural estimator of $\mathbf{W}_t$ as
\begin{equation*}
    \widehat{\mathbf{W}}_t = \mathbf{Y}_t - \widehat{\mathbf{A}} \mathbf{X}_t.
\end{equation*}

Next, we proceed to estimate $\mathcal{M}(\mathbf{Q}_1)$ and $\mathcal{M}(\mathbf{Q}_2)$. Some notations need to be defined first.
Let the $j$th column of matrices $\mathbf{X}_t, \widehat{\mathbf{W}}_t, \mathbf{W}_t, \mathbf{Q}_1, \mathbf{Q}_2$ and $\mathbf{E}_t$ be denoted as $\mathbf{x}_{t,j}, \hat{\mathbf{w}}_{t,j}, \mathbf{w}_{t,j}, \mathbf{q}_{1,j}, \mathbf{q}_{2,j}$ and $\mathbf{e}_{t,j}$, respectively.
Let $\mathbf{q}_{1,j\cdot}$ and $\mathbf{q}_{2,j \cdot}$ represent the $j$th row of $\mathbf{Q}_1$ and $\mathbf{Q}_2$, respectively. Considering the $j$th column of $\mathbf{W}_t$ and $\widehat{\mathbf{W}}_t$, we have
    \begin{align}
    \label{E5}
        \mathbf{w}_{t,j} & = \mathbf{Q}_1 \mathbf{Z}_{t} \mathbf{q}_{2,j \cdot}^{\prime} + \mathbf{e}_{t,j},\\
        \widehat{\mathbf{w}}_{t, j} & = \mathbf{A} \mathbf{x}_{t,j}  - \widehat{\mathbf{A}} \mathbf{x}_{t,j}   + \mathbf{Q}_1 \mathbf{Z}_{t} \mathbf{q}_{2,j \cdot}^{\prime} + \mathbf{e}_{t,j}  \nonumber\\
        & = (\mathbf{A}- \widehat{\mathbf{A}}) \mathbf{x}_{t,j} +  \mathbf{Q}_1 \mathbf{Z}_{t} \mathbf{q}_{2,j \cdot}^{\prime} + \mathbf{e}_{t,j}.
    \end{align}
For $i,j=1,2,...,q$, define
\begin{align}
    \label{E7}
    \boldsymbol{\Omega}_{zq,ij}(h) & = \frac{1}{T-h} \sum_{t=1}^{T-h} \text{Cov} (\mathbf{Z}_t \mathbf{q}_{2,i \cdot}^{\prime}, \mathbf{Z}_{t+h} \mathbf{q}_{2,j \cdot}^{\prime}) ,\\
    \label{E8}
    \boldsymbol{\Omega}_{w,ij}(h) & = \frac{1}{T-h} \sum_{t=1}^{T-h} \text{Cov} (\mathbf{w}_{t,i}, \mathbf{w}_{t+h, j}),
\end{align}
 By plugging \eqref{E5} into \eqref{E8} and by the assumption that the process $\{\mathbf{E}_t\}$ is white noise, it follows that
\begin{equation} \label{E9}
    \boldsymbol{\Omega}_{w,ij}(h) = \mathbf{Q}_1  \boldsymbol{\Omega}_{zq,ij}(h) \mathbf{Q}_1^{\prime},
\end{equation}
for $h \geq 1$. For a pre-determined integer $h_0$, construct the following nonnegative statistics
\begin{equation} \label{E10}
    \mathbf{M}_{1} = \sum_{h=1}^{h_0} \sum_{i=1}^{q} \sum_{j=1}^{q} \boldsymbol{\Omega}_{w,ij}(h) \boldsymbol{\Omega}_{w,ij}(h)^{\prime}.
\end{equation}
By \eqref{E9} and \eqref{E10}, it follows that
\begin{equation}
    \mathbf{M}_1 =  \mathbf{Q}_{1} \left(\sum_{h=1}^{h_0} \sum_{i=1}^{q} \sum_{j=1}^{q} \boldsymbol{\Omega}_{zq,ij}(h) \boldsymbol{\Omega}_{zq,ij}(h)^{\prime}\right) \mathbf{Q}_1^{\prime} .
 \end{equation}
 
 Suppose that the rank of matrix $\mathbf{M}$ is $k$. Since $\mathbf{M}_1$ has a $\mathbf{Q}_1$ on the left and a $\mathbf{Q}_1^{\prime}$ on the right. Hence the eigen-space of $\mathbf{M}_1$ is the 
 same as the column space of $\mathbf{Q}_1$. Consequently, we select the eigenvectors corresponding to the $k$ non-zero eigenvalues of $\mathbf{M}_1$ as the columns of $\mathbf{Q}_1$. Assuming further that $\mathbf{M}_1$ 
 has $k$ distinct positive eigenvalues and choosing the positive $\mathbf{1}^{\prime} \mathbf{q}_{1,j}$, we can uniquely define the matrix $\mathbf{Q}_1$ as follows:
 \begin{equation*}
    \mathbf{Q}_1 = (\mathbf{q}_{1,1},...,\mathbf{q}_{1,k}),
 \end{equation*}
where $\mathbf{q}_{1,i}$ is the eigenvector corresponding to the $i$th largest eigenvalues of $\mathbf{M}_1$.
 
The sample version of $\boldsymbol{\Omega}_{w,ij}(h)$ and $\mathbf{M}_1$ are
 \begin{align}
    \widehat{\boldsymbol{\Omega}}_{w,ij}(h) & = \frac{1}{T-h} \sum_{t=1}^{T-h} \hat{\mathbf{w}}_{t,i} \hat{\mathbf{w}}_{t+h, j}^{\prime}, \\
    \widehat{\mathbf{M}}_{1} & = \sum_{h=1}^{h_0} \sum_{i=1}^{q} \sum_{j=1}^{q} \widehat{\boldsymbol{\Omega}}_{w,ij}(h) \widehat{\boldsymbol{\Omega}}_{w,ij}(h)^{\prime}.
\end{align}
Then $\mathcal{M}(\mathbf{Q}_1)$ can be estimated by $\mathcal{M}(\widehat{\mathbf{Q}}_1)$, where $\widehat{\mathbf{Q}}_1 = (\hat{\mathbf{q}}_{1,1},...,\hat{\mathbf{q}}_{1,k})$, and $\hat{\mathbf{q}}_{1,i}$ represents the eigenvector of $\widehat{\mathbf{M}}_1$, 
corresponding to the $i$-th largest eigenvalues. 

 The dimensions $k$ and $r$ need to be determined before the estimation procedures. According to \citet{10.1214/12-AOS970}, let $\widehat{\lambda}_{1,1} \geq \widehat{\lambda}_{1,2} \geq \cdots \geq \widehat{\lambda}_{1,p} \geq 0$
 be the ordered eigenvalues of $\widehat{\mathbf{M}}_1$. The ratio-based estimator for $k$ is 
 \begin{equation}
    \widehat{k}=\arg \min _{1 \leq j \leq K} \frac{\widehat{\lambda}_{1,j+1}}{\widehat{\lambda}_{1,j}},
    \end{equation}
where $K$ is usually chosen to be $p/3$ or $p/2$ in practice.

For estimating the factor loading $\mathbf{Q}_2$ and $r$, we can apply the same procedure on the transpose of $\widehat{\mathbf{W}}_t$'s to construct $\widehat{\mathbf{M}}_2$. 
Further the estimation of normalized latent 
factor process $\mathbf{Z}_t$ is
\begin{equation*}
    \widehat{\mathbf{Z}}_t = \widehat{\mathbf{Q}}_1^{\prime} \widehat{\mathbf{W}}_t \widehat{\mathbf{Q}}_2.
\end{equation*}
At last, the factor signal part $ \mathbf{L}_t =  \mathbf{R} \mathbf{F}_t  \mathbf{C}^{\prime}$ and the total signal part $\mathbf{S}_t = \mathbf{A}\mathbf{X}_t + \mathbf{R} \mathbf{F}_t  \mathbf{C}^{\prime}$ can be estimated by
 \begin{equation*}
 \begin{aligned}
     \widehat{\mathbf{L}}_t & =   \widehat{\mathbf{Q}}_1 \widehat{\mathbf{Q}}_1^{\prime} \widehat{\mathbf{W}}_t \widehat{\mathbf{Q}}_2 \widehat{\mathbf{Q}}_2^{\prime},\\
     \widehat{\mathbf{S}}_t & = \widehat{\mathbf{A}} \mathbf{X}_t + \widehat{\mathbf{Q}}_1 \widehat{\mathbf{Q}}_1^{\prime} \widehat{\mathbf{W}}_t \widehat{\mathbf{Q}}_2 \widehat{\mathbf{Q}}_2^{\prime},
 \end{aligned} 
 \end{equation*}
respectively.
\section{Theoretical Properties} \label{Section 4}
\noindent{}The asymptotic properties are built based on $T$, $p$ and $q$ tend to infinity, while keep $k$, $r$ and $m$ fix.
Define 
\begin{equation*}
    \boldsymbol{\Sigma}_f(h) = \frac{1}{T-h} \text{Cov} \left(\text{vec}(\mathbf{F}_t), \text{vec}(\mathbf{F}_{t+h})\right)\qquad \text{and}\qquad \boldsymbol{\Sigma}_{e} = \text{Var}(\text{vec}(\mathbf{E}_t)). 
\end{equation*}

\noindent {\bf Condition 1.} 
    The process $\{ (\text{vec}(\mathbf{Y}_t), \text{vec}(\mathbf{X}_t), \text{vec}(\mathbf{F}_t))\}$ is $\alpha$-mixing with the mixing coefficients 
satisfying the condition $\sum_{k=1}^{\infty} \alpha(k)^{1-2/\gamma} < \infty$ for some $\gamma > 2$, where
\begin{equation*}
    \alpha(k) =  \mathop{\sup}_{i} \mathop{\sup}_{A \in \mathcal{F}_{-\infty }^{i},\ B \in \mathcal{F}_{i+k}^{\infty}} | P(A) P(B) - P(AB) |, 
\end{equation*} 
and $\mathcal{F}_{i }^{j}$ is the $\sigma$-field generated by $\{(\text{vec}(\mathbf{Y}_t), \text{vec}(\mathbf{X}_t), \text{vec}(\mathbf{F}_t)) :\ i \leq t\leq j\}$. 

\vspace{.1in}

\noindent {\bf Condition 2.} 
    For any $i=1,...,k$, $j=1,...,r$, $l=1,...,m$, $u=1,...,p$, $g=1,...,q$ and $t=1,...,T$, we assume $E(|f_{t,ij}|^{2\gamma}) \leq C$, $E(|x_{t,lg}|^{2\gamma}) \leq C$ and 
$E(|e_{t,ug}|^{2\gamma}) \leq C$, where $C$ is a constant, $\gamma$ is given in Condition 1, and $f_{t,ij}$ is the $(i,j)$th element of $\mathbf{F}_t$, $x_{t,lg}$ is the $(l,g)$th element 
of $\mathbf{X}_t$, $e_{t,ug}$ is the $(u,g)$th element of $\mathbf{E}_t$.
\vspace{.1in}

\noindent {\bf Condition 3.} 
    There exists an integer $h$ satisfying $ 1 \leq h\leq h_0$ such that the rank of $\boldsymbol{\Sigma}_f(h)$ is $k^{*}=\max(k, r)$ and $\|\boldsymbol{\Sigma}_f(h)\|_2 \asymp O(1) \asymp \sigma_{k^*}(\boldsymbol{\Sigma}_f(h))$. 
For $i= 1,...,k$ and $ j =1,...,r$, $\frac{1}{T-h} \sum_{t=1}^{T-h} \text{Cov}(F_{t,i\cdot}, F_{t+h, i\cdot}) \neq \mathbf{0}$ and $\frac{1}{T-h} \sum_{t=1}^{T-h} \text{Cov}(F_{t,\cdot j }, F_{t+h, \cdot j }) \neq \mathbf{0}$ .
\vspace{.1in}

\noindent {\bf Condition 4.} 
    $E(\mathbf{X}_t \mathbf{X}_t^T) = \mathbf{P}_t$, a positive definite matrix with $(1/T) \sum_{t=1}^{T} \mathbf{P}_t \rightarrow \mathbf{P}$,  a positive definite matrix. And $\lambda_{\min}(\mathbf{P})\asymp q$.
\vspace{.1in}

\noindent {\bf Condition 5.} 
    All the components of matrix $\boldsymbol{\Sigma}_e$ are bounded as $p$ and $q$ increase to infinity.
\vspace{.1in}

\noindent {\bf Condition 6.} 
    There exists two constants $\delta_1$ and $\delta_2$ $\in [0,1]$ such that $\|\mathbf{R}\|_2^2 \asymp p^{1-\delta_1} \asymp \|\mathbf{R}\|_{\min}^2$ and $\|\mathbf{C}\|_2^2 
\asymp q^{1-\delta_2} \asymp \|\mathbf{C} \|_{\min}^2$.

Condition 4 allows the expectation of row covariance matrix of $\{\mathbf{X}_t\}$ might be different for different $t$, so long as the limit of $(1/T) \sum_{t=1}^{T} E(\mathbf{X}_t \mathbf{X}_t^T)$ can be consistently estimated by $(1/T) \sum_{t=1}^{T} (\mathbf{X}_t \mathbf{X}_t^T)$.

The parameters $\delta_1$ and $\delta_2$ are referred to as the strength of row factors and the strength of column factors in Condition 6, respectively. If $\delta_i =0$, the corresponding factors are called strong factors. If $1> \delta_i >0$, 
the corresponding factors are called weak factors. The smaller the $\delta$'s, the stronger the factors.
\vspace{.1in}

\noindent {\bf Condition 7.} 
    $\mathbf{M}_1$ and $\mathbf{M}_2$ have $k$ and $r$ distinct positive eigenvalues, respectively.

\begin{theorem} \label{Theorem 1}
    As $T \rightarrow \infty$ and $ p, q \rightarrow \infty$, it holds that
    \begin{equation*}
        \| \widehat{\mathbf{A}} -  \mathbf{A} \|_F  = O_p( p^{1/2}  T^{-1/2}).
    \end{equation*}
\end{theorem}
\autoref{Theorem 1} establishes the convergence rate of the estimator for the coefficient matrix $\mathbf{A}$. 
Because the number of parameters contained in matrix $\mathbf{A}$ is $p\times m$, it is reasonable that the convergence speed depends on $p$. In particular, when $p/T$ approaches zero,  
$\widehat{\mathbf{A}}$  is a consistent estimator for $\mathbf{A}$.
\begin{theorem} \label{Theorem 2}
    Under Conditions 1-7 and $p^{\delta_1} q^{\delta_2} T^{-1/2} = o(1)$, it holds that 
    \begin{equation*}
        \| \widehat{\mathbf{Q}}_{i} - \mathbf{Q}_{i} \|_2 = O_p\left(p^{\delta_1} q^{\delta_2} T^{-1/2} \right), \quad for\ i=1,2.
    \end{equation*}
\end{theorem}
The convergence rates state in \autoref{Theorem 2} are exactly the same as in \citet{wang2019factor} of the Theorem 1. That means adding the known factors term $\mathbf{A} \mathbf{X}_t$ does not alter the convergence rates of the factor loadings. Both the 
strength of row factors and strength of column factor impact the convergence rate.
\begin{theorem} \label{Theorem 3}
Under Conditions 1-7 and $p^{\delta_1} q^{\delta_2} T^{-1 / 2}=o(1)$, the eigenvalues $\left\{\widehat{\lambda}_{1, 1}, \ldots, \widehat{\lambda}_{1, p}\right\}$ of $\widehat{\mathbf{M}}_1$ which are sorted in descending order satisfy
$$
\text { and } \quad \begin{aligned}
\left|\widehat{\lambda}_{1, j}-\lambda_{1, j}\right| & =O_p\left( p^{2-\delta_1} q^{2-\delta_2} T^{-1 / 2}\right), \quad \text { for } j=1,2, \ldots, k, \\
\left|\widehat{\lambda}_{1, j}\right| & =O_p\left(p^2 q^2 T^{-1}\right), \quad \text { for } j=k+1, \ldots, p,
\end{aligned}
$$
where $\lambda_{1, 1}>\lambda_{1, 2} \ldots>\lambda_{1, k}$ are eigenvalues of $\mathbf{M}_1$. And the eigenvalues $\left\{\widehat{\lambda}_{2, 1}, \ldots, \widehat{\lambda}_{2, q}\right\}$ of $\widehat{\mathbf{M}}_2$ which are sorted in descending order satisfy
$$
\text { and } \quad \begin{aligned}
\left|\widehat{\lambda}_{2, j}-\lambda_{2, j}\right| & =O_p\left( p^{2-\delta_1} q^{2-\delta_2} T^{-1 / 2}\right), \quad \text { for } j=1,2, \ldots, r, \\
\left|\widehat{\lambda}_{2, j}\right| & =O_p\left(p^2 q^2 T^{-1}\right), \quad \text { for } j=r+1, \ldots, q,
\end{aligned}
$$
where $\lambda_{2, 1}>\lambda_{2, 2} \ldots>\lambda_{2, r}$ are eigenvalues of $\mathbf{M}_2$. 
\end{theorem}

Under the condition $p^{\delta_1} q^{\delta_2} T^{-1/2}=o(1)$, the zero eigenvalues of $\widehat{\mathbf{M}}_i$ for $i=1,2$ gives the faster convergence rate of the nonzero eigenvalues. This provides the theoretical support for the 
the ratio-based estimator introduced in Section 3.

\begin{theorem} \label{Theorem 4}
    Under Conditions 1-7, $p^{\delta_1} q^{\delta_2} T^{-1/2} = o(1)$ and $\|\boldsymbol{\Sigma}_e\|_2$ is bounded, it holds that 
    \begin{equation*}
       p^{-1/2} q^{-1/2} \| \widehat{\mathbf{S}}_{t} - \mathbf{S}_{t} \|_2 = O_p(p^{\delta_1/2} q^{\delta_2/2} T^{-1/2} + p^{-1/2}q^{-1/2}).
    \end{equation*}
\end{theorem} 

\autoref{Theorem 4} shows that the total signal $\mathbf{S}_t$ can  be consistently estimated  only as the dimensions of the matrix time series $\mathbf{Y}_t$ tend to infinity.

To evaluate the distance between the estimated loading space  and true loading space, we adopt the distance measure used in \citet{chang2015high} and \citet{liu2016regime}:

For two orthogonal matrices $\mathbf{O}_1$ and $\mathbf{O}_2$ of size $ p \times q_1$ and $p \times q_2$, define 
\begin{equation}
    \mathcal{D}\left(\mathbf{O}_1, \mathbf{O}_2\right)=\left(1-\frac{1}{\max \left(q_1, q_2\right)} \operatorname{Tr}\left(\mathbf{O}_1 \mathbf{O}_1^{\prime} \mathbf{O}_2 \mathbf{O}_2^{\prime}\right)\right)^{1 / 2}.
\end{equation}
It a quality between 0 and 1. If the column spaces of $\mathbf{O}_1$ and $\mathbf{O}_2$ are the same, the value is equal to 0. The column spaces of $\mathbf{O}_1$ and $\mathbf{O}_2$  are orthogonal if it is equal to 1.

\begin{theorem} \label{Theorem 5}
    Under Conditions 1-7 and $p^{\delta_1} q^{\delta_2} T^{-1/2} = o(1)$, it holds that 
    \begin{equation*}
       \mathcal{D}\left(\widehat{\mathbf{Q}}_i, \mathbf{Q}_i \right) = O_p\left(p^{\delta_1} q^{\delta_2} T^{-1/2} \right), \quad for\ i=1,2.
    \end{equation*}
\end{theorem}

\section{Simulation}
\noindent{}In this section, we show the performance of the estimation procedures proposed in Section 3. 
The observed data $\mathbf{Y}_t$ are generated from the model \eqref{E1},  
\begin{equation} 
    \mathbf{Y}_t = \mathbf{A} \mathbf{X}_t+ \mathbf{R} \mathbf{F}_t \mathbf{C}^{\prime} +\mathbf{E}_t,\qquad t=1,2,...,T.
 \end{equation}
 We set the dimensions of $\mathbf{Y}_t$ at $(p,\ q)=(10,\ 10),\ (20,\ 20),\ (20,\ 50),\ (50,\ 50)$. The sample size $T$ is chosen as 
 $0.5pq,\ pq,\ 2pq$. Let each column of $\mathbf{X}_t$ follows the VAR(1) model:
 \begin{equation*}
    \mathbf{x}_{t,j} = \begin{pmatrix}
        5/8 & 1/8 & 1/8 \\
        1/8 & 5/8 & 1/8 \\
        1/8 & 1/8 & 5/8
    \end{pmatrix} \mathbf{x}_{t-1,j} + \boldsymbol{\epsilon}_{t,j},
 \end{equation*} where $\boldsymbol{\epsilon}_{t,j}\ \sim\ N(\mathbf{0}, \mathbf{I}_3),\ j=1,...,q$. The entries of matrix $\mathbf{A}$ are 
 independently sampled from the uniform distribution $U(-1,\ 1)$ . We choose the dimensions of the latent factors process $\mathbf{F}_t$ to be $k=3$ and $r=3$. The entries of $\mathbf{F}_t$ follow $kr$ independent AR(1) processes with 
 Gaussian white noise $\mathcal{N}(0,\ 1)$ innovations, $\text{vec}(\mathbf{F}_t) = \boldsymbol{\Phi}_F \text{vec}(\mathbf{F}_{t-1}) + \boldsymbol{\varepsilon}_t
$ with $\boldsymbol{\Phi}_F$ being the diagonal matrix with entries $(-0.5, 0.6, 0.5, 0.8, -0.4, 0.6, 0.7, 0.3, 0.4)$. Let the factor loadings $ \mathbf{R}$ and $\mathbf{C}$ with the elements drawn from the 
uniform distribution on $(-p^{-\delta_1/2},\ p^{-\delta_1/2})$ and $(-q^{-\delta_2/2},\ q^{-\delta_2/2})$, respectively. Three combinations of $(\delta_1,\ \delta_2)$ are considered in our simulation: 
(0.5,\ 0.5), (0.5, 0), (0, 0). The error term $\mathbf{E}_t$ is a white-noise process from a matrix normal distribution $ \mathcal{MN}(\mathbf{0},\ \boldsymbol{\Sigma}_1,\ \boldsymbol{\Sigma}_2)$, that is  $\text{vec}(\mathbf{E}_t) \sim 
\mathcal{N}(\mathbf{0},\ \boldsymbol{\Sigma}_2 \otimes \boldsymbol{\Sigma}_1)$. Let $\boldsymbol{\Sigma}_1$ and $\boldsymbol{\Sigma}_2$ be matrices with 1 on the diagonal, and the off-diagonal entries are 0.2. All the simulation results are presented under $h_0=1$, since the results are similar with small $h_0$.   

Table \ref{first result} reports the estimation performance of the coefficient matrix $\mathbf{A}$. The results show that increasing the strength of row factors or the strength of column factors do not significantly alter the estimation accuracies. And extending the temporal span of the data can improve the estimation accuracies.

 \begin{table}[h!]
 \centering
    \tabcolsep  = 0.15cm
        \caption{Means  of $ \| \mathbf{A} - \widehat{\mathbf{A}} \|_F/p^{1/2}$  over 200 simulation runs.}
        \begin{threeparttable}
    \begin{tabular}{ccccccccccc}
        \hline 
        \multicolumn{5}{c}{} & \multicolumn{1}{c}{$T = 0.5*p*q $} & \multicolumn{1}{c}{$ T = p*q $} &\multicolumn{1}{c}{$ T= 2*p*q $}\\
        \hline\rule{0pt}{15pt}
        $ \delta_1 $ & $ \delta_2 $ & $ p $ & $ q $ & $m$ & $ \| \mathbf{A} - \widehat{\mathbf{A}} \|_F/p^{1/2} $  & $ \| \mathbf{A} - \widehat{\mathbf{A}} \|_F/p^{1/2} $  & $\| \mathbf{A} - \widehat{\mathbf{A}} \|_F/p^{1/2}$  \\
        \hline
        0.5 & 0.5 & 10 & 10 & 5 & 0.101 &0.073 &0.050 \\
        && 20 & 20 & 5 & 0.034 &0.024&0.017\\
        && 20 & 50 & 5 & 0.013 & 0.009 & 0.007 \\
        && 50 & 50 & 5 & 0.008 & 0.006 & 0.004 \\
        \hline
        0.5 & 0 & 10 & 10 & 5 & 0.111 &0.082 &0.058 \\
        && 20 & 20 & 5 & 0.038 & 0.027 & 0.019 \\
        && 20 & 50 & 5 & 0.016 & 0.011 & 0.007  \\
        && 50 & 50 & 5 & 0.009 & 0.007 & 0.005 \\
        \hline
        0 & 0 & 10 & 10 & 5 &  0.155 & 0.113 &0.083 \\
        && 20 & 20 & 5 & 0.051 & 0.038 & 0.028 \\
        && 20 & 50 & 5 & 0.022 & 0.016 & 0.011 \\
        && 50 & 50 & 5 & 0.013 & 0.009 & 0.007\\
        \hline 
    \end{tabular}
     \begin{tablenotes}[para,flushleft]
		\footnotesize
		\item Note:  $\delta_1,\ \delta_2$ 
		 represent the strength for the  row factors and the column factors, respectively.
	  \end{tablenotes} 
   \end{threeparttable}
     \label{first result}
\end{table} 

We next investigate the accuracies of the estimated column factor loading space and row factor loading space. Assuming that the dimensions of latent factors $k \times r$ are known, we then calculate $\mathcal{D}(\widehat{\mathbf{Q}}_1, \mathbf{Q}_1)$ and $\mathcal{D}(\widehat{\mathbf{Q}}_2, \mathbf{Q}_2)$, which are shown in Table \ref{Table 2}. The results suggest that when both row factors and column factors are weak, the estimation performance is subpar. Additionally, the estimation accuracies improve with increases in sample size or the strength of the row or column loading matrices.
 \begin{table}[h!]
    \tabcolsep  = 0.15cm
        \caption{Means and standard deviations  (in parentheses) of $ \mathcal{D}(\widehat{\mathbf{Q}}_{1}, \mathbf{Q}_{1}) $ and $ \mathcal{D}(\widehat{\mathbf{Q}}_{2}, \mathbf{Q}_{2})$  over 200 simulation runs. For ease of presentation, all numbers in this table are the true numbers multipied by 10.}
        \begin{threeparttable}
    \begin{tabular}{ccccccccccc}
        \hline 
        \multicolumn{5}{c}{} & \multicolumn{2}{c}{$T = 0.5*p*q $} & \multicolumn{2}{c}{$ T = p*q $} &\multicolumn{2}{c}{$ T= 2*p*q $}\\
        \hline\rule{0pt}{15pt}
        $ \delta_1 $ & $ \delta_2 $ & $ p $ & $ q $ & $m$ & $ \mathcal{D}(\widehat{\mathbf{Q}}_{1}, \mathbf{Q}_{1}) $ & $ \mathcal{D}(\widehat{\mathbf{Q}}_{2}, \mathbf{Q}_{2}) $ & $ \mathcal{D}(\widehat{\mathbf{Q}}_{1}, \mathbf{Q}_{1}) $ & $ \mathcal{D}(\widehat{\mathbf{Q}}_{2}, \mathbf{Q}_{2}) $ & $ \mathcal{D}(\widehat{\mathbf{Q}}_{1}, \mathbf{Q}_{1}) $ & $ \mathcal{D}(\widehat{\mathbf{Q}}_{2}, \mathbf{Q}_{2}) $ \\
        \hline
        0.5 & 0.5 & 10 & 10 & 5 & 4.10(0.38) &5.74(0.55) &4.37(0.99)&4.73(0.71)&4.19(0.57)&5.68(0.16) \\
        && 20 & 20 & 5 & 5.48(0.18)&5.62(0.19)&5.59(0.28)&5.75(0.11)&5.75(0.07)&4.62(0.14)\\
        && 20 & 50 & 5 &5.68(0.15)&5.84(0.05)&5.55(0.07)&5.76(0.02)&4.16(0.05)&5.72(0.03)\\
        && 50 & 50 & 5 & 5.82(0.04)&5.59(0.04)&5.67(0.02)&5.57(0.04)&5.58(0.06)&5.77(0.01)\\
        \hline
        0.5 & 0 & 10 & 10 & 5 & 5.87(0.25)&4.06(0.46) &4.23(0.59)&4.08(0.68)&4.96(0.42)&1.49(0.83) \\
        && 20 & 20 & 5 & 4.65(0.91)&4.39(0.69)&2.43(0.37)&2.32(0.41)&0.65(0.12)&0.76(0.21)\\
        && 20 & 50 & 5 & 1.14(0.35)&1.09(0.24)&0.74(0.15)&0.63(0.15)&0.58(0.12)&0.52(0.09)\\
        && 50 & 50 & 5 & 1.55(0.24)&0.88(0.17)&0.32(0.10)&0.28(0.05)&0.32(0.05)&0.35(0.05)\\
        \hline
        0 & 0 & 10 & 10 & 5 &  1.56(0.69)& 3.38(1.22) &0.78(0.19)&0.94(0.33)&0.69(0.24)&0.59(0.16) \\
        && 20 & 20 & 5 & 0.54(0.10)&0.69(0.23)&0.48(0.12)&0.40(0.10)&0.20(0.03)&0.29(0.07)\\
        && 20 & 50 & 5 & 0.19(0.03)&0.28(0.04)&0.18(0.04)&0.20(0.02)&0.08(0.01)&0.13(0.01)\\
        && 50 & 50 & 5 & 0.13(0.02)&0.13(0.02)&0.09(0.01)&0.09(0.01)&0.05(0.01)&0.05(0.01)\\
        \hline 
        \end{tabular}
        \begin{tablenotes}[para,flushleft]
		\footnotesize
		\item Note:  $\delta_1,\ \delta_2$ 
		 represent the strength for the  row factors and the column factors, respectively.
	  \end{tablenotes} 
   \end{threeparttable}
 \label{Table 2}
\end{table} 

Table \ref{Table 3} reports the relative frequencies of correctly estimating the dimensions of factors. It is noteworthy that when the column or row factors are weak, the estimation method struggles to accurately estimate the dimensions of latent factors. When both row and column factors are strong, the relative frequency of correctly estimating the dimensions of factors improves with increasing sample size.

\begin{table}[h!]
\centering
    \tabcolsep  = 0.15cm
        \caption{Relative frequencies of correctly estimating the dimensions of factors.}
      \begin{threeparttable}  
    \begin{tabular}{ccccccccccc}
        \hline 
        \multicolumn{5}{c}{} & \multicolumn{2}{c}{$T = 0.5*p*q $} & \multicolumn{2}{c}{$ T = p*q $} &\multicolumn{2}{c}{$ T= 2*p*q $}\\
        \hline\rule{0pt}{15pt}
        $ \delta_1 $ & $ \delta_2 $ & $ p $ & $ q $ & $m$ & \multicolumn{2}{c}{$f $}& \multicolumn{2}{c}{$f $}&\multicolumn{2}{c}{$ f $} \\
        \hline
        0.5 & 0.5 & 10 & 10 & 5 & \multicolumn{2}{c}{0.0}& \multicolumn{2}{c}{0.0}&\multicolumn{2}{c}{0.005} \\
        && 20 & 20 & 5 & \multicolumn{2}{c}{0.0}& \multicolumn{2}{c}{0.0}&\multicolumn{2}{c}{0.0}\\
        && 20 & 50 & 5 &\multicolumn{2}{c}{0.0}& \multicolumn{2}{c}{0.0 }&\multicolumn{2}{c}{0.0}\\
        && 50 & 50 & 5 & \multicolumn{2}{c}{0.0}& \multicolumn{2}{c}{0.0}&\multicolumn{2}{c}{0.0}\\
        \hline
        0.5 & 0 & 10 & 10 & 5 & \multicolumn{2}{c}{0.0}& \multicolumn{2}{c}{0.08}&\multicolumn{2}{c}{0.0} \\
        && 20 & 20 & 5 & \multicolumn{2}{c}{0.0}& \multicolumn{2}{c}{0.0}&\multicolumn{2}{c}{0.0}\\
        && 20 & 50 & 5 & \multicolumn{2}{c}{0.0}& \multicolumn{2}{c}{0.0}&\multicolumn{2}{c}{0.0}\\
        && 50 & 50 & 5 & \multicolumn{2}{c}{0.0}& \multicolumn{2}{c}{0.0}&\multicolumn{2}{c}{0.0}\\
        \hline
        0 & 0 & 10 & 10 & 5 &  \multicolumn{2}{c}{0.0}& \multicolumn{2}{c}{0.475}&\multicolumn{2}{c}{0.15} \\
        && 20 & 20 & 5 & \multicolumn{2}{c}{0.07}& \multicolumn{2}{c}{0.765}&\multicolumn{2}{c}{0.555}\\
        && 20 & 50 & 5 & \multicolumn{2}{c}{1.0}& \multicolumn{2}{c}{1.0}&\multicolumn{2}{c}{1.0}\\
        && 50 & 50 & 5 & \multicolumn{2}{c}{1.0}& \multicolumn{2}{c}{1.0}&\multicolumn{2}{c}{1.0}\\
        \hline 
    \end{tabular}
    \begin{tablenotes}[para,flushleft]
		\footnotesize
		\item Note:  $\delta_1,\ \delta_2$ 
		 represent the strength for the  row factors and the column factors, respectively.
	  \end{tablenotes} 
   \end{threeparttable}
    \label{Table 3}
\end{table}

Finally, we investigate the performance of recovering the signal part $\mathbf{S}_t$. 
We  evaluate the performance of recovering by the normalized spectral norm, that is
\begin{equation}
    \mathcal{D}(\widehat{\mathbf{S}}, \mathbf{S} )= p^{-1/2} q^{-1/2} \frac{\sum_{t=1}^T \| \widehat{\mathbf{S}}_t - \mathbf{S}_t\|_2 }{T} .
\end{equation}
We take $\delta_1 = \delta_2 = 0$, $p = q= 10,20,50$, and $T=50, 200, 1000,5000$. Table \ref{tab:label4} shows the results of $\mathcal{D}(\widehat{\mathbf{S}}, \mathbf{S} )$. We observe that the recovery performance improves with increasing sample dimensions or sample size. When the sample dimensions are fixed, the recovery performance ceases to improve after a certain point as the sample size increases.

\begin{table}[h!]
    \centering
    \caption{Means and standard deviations  (in parentheses) of $ \mathcal{D}(\widehat{\mathbf{S}}, \mathbf{S}) $   over 200 simulation runs. For ease of presentation, all numbers in this table are the true numbers multipied by 10.}
      \begin{tabular}{c|cccc}
\hline \multirow{2}{*}{$p=q, m=5$} & \multicolumn{4}{c}{$\mathcal{D}(\widehat{\mathbf{S}}, \mathbf{S})$} \\
\cline { 2 - 5 } & $T=50$ & $T=200$ & $T=1000$ & $T=5000$ \\
\hline 10 & $3.99(0.04)$ & $3.12(0.00)$ & $2.95(0.00)$ & $3.02(0.00)$ \\
20 & $2.03(0.02)$ & $1.44(0.00)$ & $1.43(0.00)$ & $1.21(0.00)$ \\
50 & $1.36(0.02)$ & $0.83(0.00)$ & $0.50(0.00)$ & $0.55(0.00)$ \\
\hline
\end{tabular}
    \label{tab:label4}
\end{table}

\section{Real Data Analysis}
\begin{figure}
    \centering
    \includegraphics[width=0.8\textwidth]{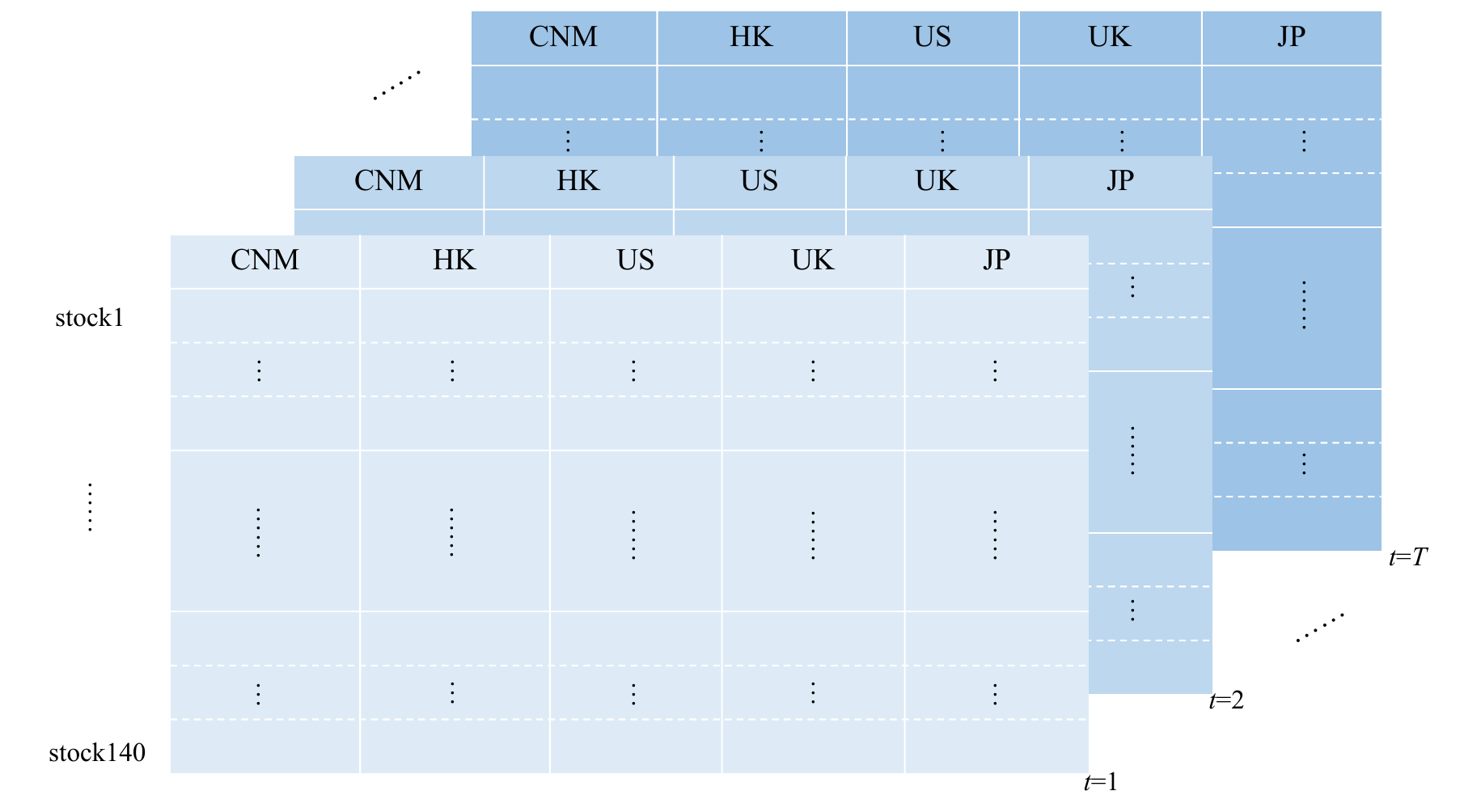}
    \caption{Data structure of multi-region and multi-industry stock returns.}
    \label{Data structure}
\end{figure}

\begin{figure}
    \centering
    \includegraphics[width=1\textwidth]{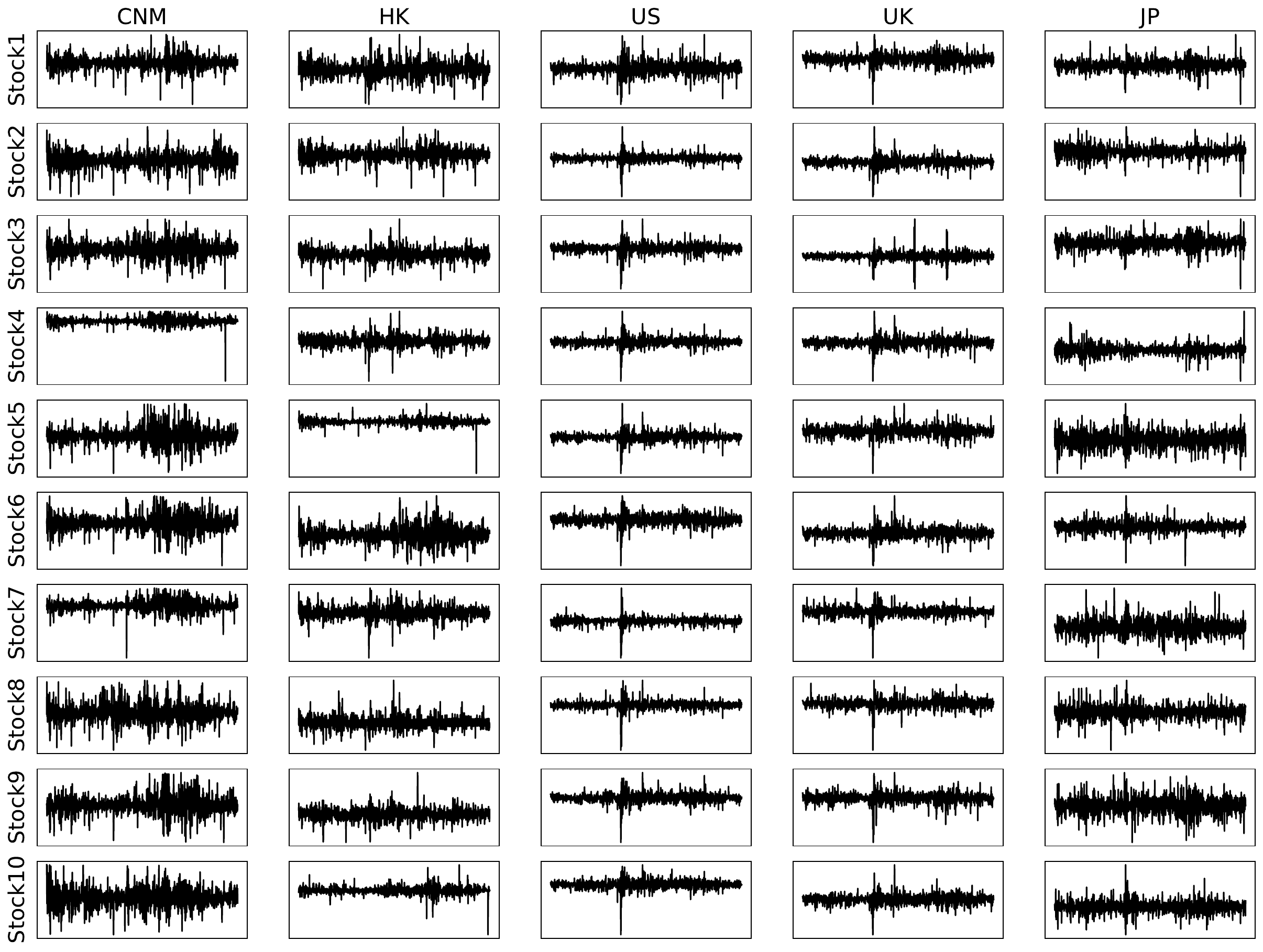}
    \caption{Time series of five regions of daily returns in Energy Industry.}
    \label{Energy Industry}
\end{figure}

\noindent{}In this section, we illustrate our methodology through the modeling of daily stocks' returns for 5 regions, including Chinese Mainland (CNM), Hong Kong (HK), the United States (US), the United Kingdom (UK), and Japan (JP). The stocks are collected from the Wind database, covering a period of 1435 trading days ($T = 1435$) from January 2018 to November 2023. According to the Wind industry classification standard, stocks in each region span 14 industries, namely Energy, Materials, Industrials, Automobiles and Auto Parts, Durables and Apparel, Consumer Service, Media, Retail, Consumer Staples, Healthcare, Financials, Information Technology, Utilities, and Real Estate. Our stock selection process within each industry involved the exclusion of stocks with incomplete data, retaining the top 10 stocks based on their total market capitalization. Daily returns are computed based on closing prices, then 
 we apply the standardization to the returns data. The data structure is visualized in Figure \ref{Data structure}. The visualization results, exemplified using Energy Industry in five regions, are presented in Figure \ref{Energy Industry}.

\begin{figure}
    \centering
    \includegraphics[width=1\textwidth]{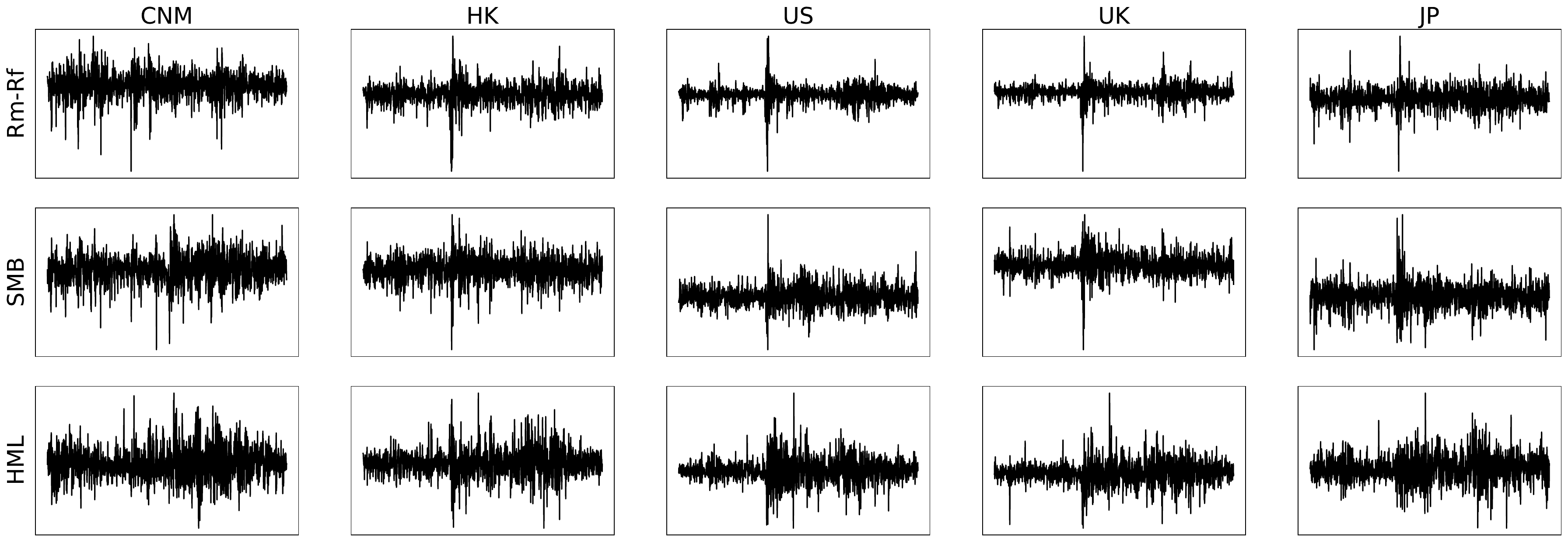}
    \caption{Time series of five regions of Fama-French three factors.}
    \label{three factors}
\end{figure}

Subsequently, we collect Fama-French three factors as known factors $\mathbf{X}_t$ in our model. These three factors include the Market Risk Premium Factor (Rm-Rf), Company Size Factor (SMB), and Book-to-Market Ratio Factor (HML), derived from the Fama-French three-factor model proposed by \citet{fama1993common}. This model is utilized for explaining and predicting the expected returns of stocks and investment portfolios, currently widely applied in academic research and investment practices (e.g., \citet{muhammad2020fama}, \cite{rohuma2023fama}). Therefore, it is reasonable to take Fama-French three factors as  known factors $\mathbf{X}_t$ in our model. Three-factor data for the Chinese mainland market are collect from the Central University of Finance and Economics School of Finance website, while data for other markets are obtained from the Fama/French Data Library. Due to data unavailability, we use Fama/French Asian market data (including the Hong Kong market) as a proxy for the Hong Kong market and European market data (including the Great Britain market) as a proxy for the UK market. After forward-filling missing values in each original time series and standardizing the data, visualizations of the three factors across the five regions are presented in Figure \ref{three factors}.

We analyze the estimated regression coefficient matrix $\widehat{\mathbf{A}}$ first. $\widehat{\mathbf{A}}$ is a $140\times 3$ matrix with each row representing the impact of three factors on the return of corresponding stock. We have selected the first 20 rows of the estimated matrix $\widehat{\mathbf{A}}$ in Table \ref{A_hat} for illustration purposes, while the remaining rows exhibit similar patterns. The coefficients corresponding to Rm-Rf are consistently in the range of 0.4 to 0.6, all positive, indicating a clear positive correlation between stock returns and Rm-Rf. The coefficients corresponding to SMB are within the range of -0.1 to 0.1, suggesting a relatively weak correlation between stock returns and SMB. Similarly, there is a positive correlation between stock returns and HML, but it is relatively weaker compared to the correlation with Rm-Rf.
\begin{table}[!htbp]
    \centering
    \caption{The first twenty rows of matrix $\widehat{\mathbf{A}}$ (Transposed).}
    \begin{tabular}{lccccccccccc}
    \toprule
     & S1 & S2 & S3 & S4 & S5 & S6 & S7 & S8 & S9 & S10 \\
    \midrule
    Rm-Rf & 0.46 & 0.51 & 0.47 & 0.49 & 0.47 & 0.50 & 0.50 & 0.46 & 0.50 & 0.45 \\
    SMB & 0.00 & -0.01 & -0.02 & -0.01 & 0.08 & 0.07 & 0.09 & 0.10 & 0.09 & 0.10 \\
    HML & 0.36 & 0.37 & 0.32 & 0.35 & 0.35 & 0.34 & 0.26 & 0.26 & 0.31 & 0.29 \\
    \midrule
    & S11 & S12 & S13 & S14 & S15 & S16 & S17 & S18 & S19 & S20 \\
    \midrule
    Rm-Rf & 0.53 & 0.44 & 0.53 & 0.57 & 0.55 & 0.57 & 0.53 & 0.44 & 0.40 & 0.55 \\
    SMB & -0.03 & -0.02 & 0.00 & -0.03 & 0.06 & 0.01 & 0.03 & 0.08 & 0.04 & 0.02 \\
    HML & 0.16 & 0.00 & 0.19 & 0.22 & 0.24 & 0.16 & 0.19 & 0.15 & 0.00 & 0.15 \\
    \bottomrule
    \end{tabular}
\label{A_hat}
\end{table}

\begin{figure}
    \centering
    \includegraphics[width=1\textwidth]{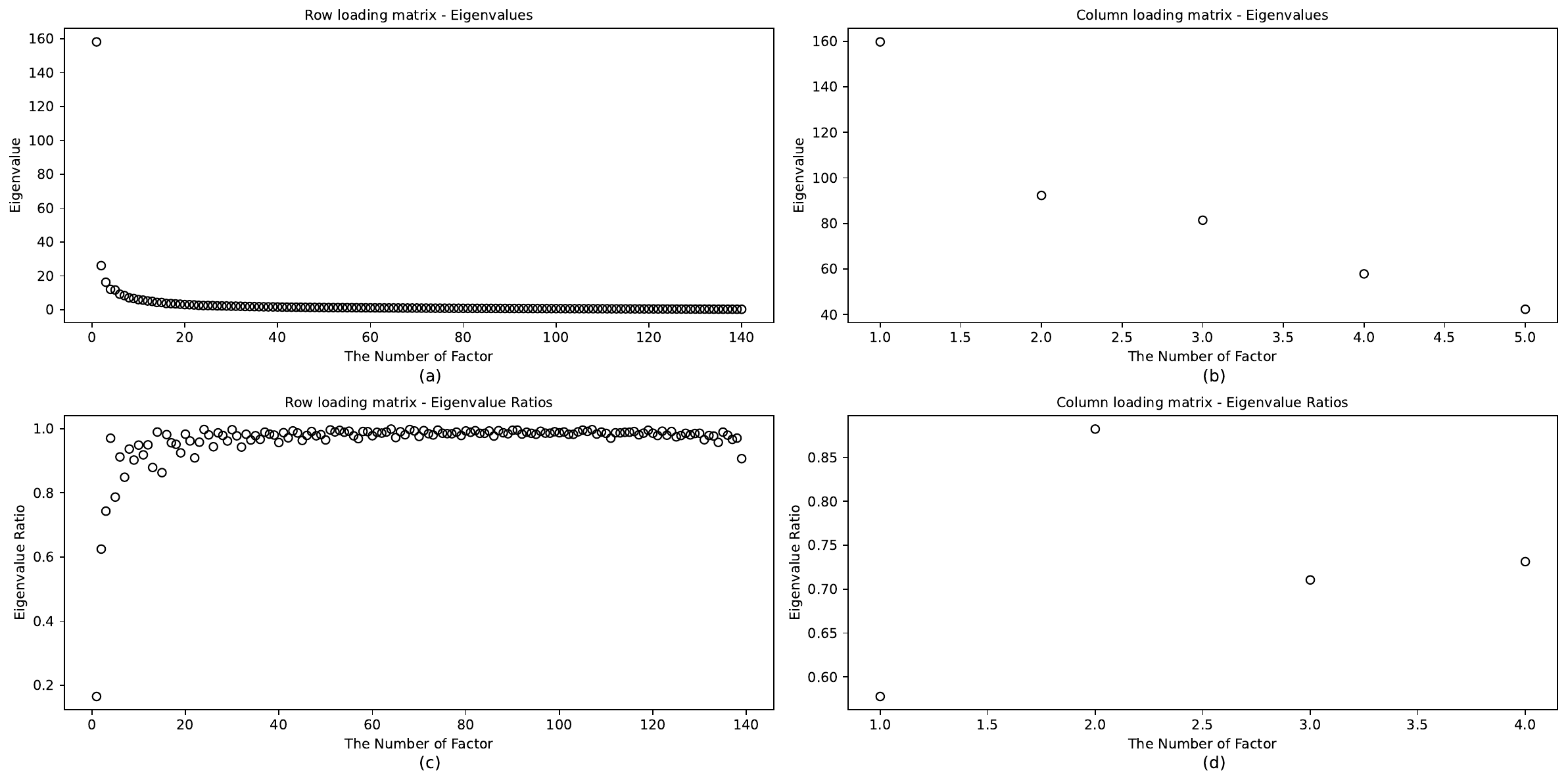}
    \caption{The eigenvalues and their ratios of $\widehat{\mathbf{M}}_1$ and $\widehat{\mathbf{M}}_2$ for row factors and column factors.}
    \label{The_eigenvalues}
\end{figure}

Figure \ref{The_eigenvalues}  shows the eigenvalues and their ratios of $\widehat{\mathbf{M}}_1$ and $\widehat{\mathbf{M}}_2$ for row factors and column factors. Although the estimated dimensions $k$ and $r$ are both 1, we use $k=2$ and $r=1$ here for accessible illustration. Estimation is done using $h_0=1$.

Next, we will provide a brief analysis of the estimated row and column factor loading matrices. The estimated row loading matrix undergoes rotation to maximize its variance, revealing that the stocks are grouped into 4 groups. Group 1, 2, 3, and 4 load heavily on the first row, load heavily on second row, load lightly on both rows, and load heavily on both rows, respectively. Table \ref{Group structure} shows the grouping result of stocks from various industries. We discover that stocks from the same industry tend to be grouped together in the same category. For example, the stocks from the Energy, Automobiles and Auto Parts and Consumer Service sectors are mainly grouped in Group 1, and the stocks from the Consumer Staples, Healthcare and Utilities sectors are mainly grouped in Group 2. Such grouping results align with our expectations. Table \ref{Column loading matrix} presents the estimated column loading matrix in our model, while Table \ref{Column loading matrix2} displays the estimated column loading matrix in the matrix factor model. The results indicate that, in the presence of known factors, the impact of latent column factor on the returns of US stocks can be disregarded.
\begin{table}[!htbp]
    \centering
    \caption{Grouping of stocks.}
\begin{tabular}{lcccccc}
\toprule Group & \multicolumn{1}{l}{1} & \multicolumn{1}{l}{2} & \multicolumn{1}{l}{3} & \multicolumn{1}{l}{4} \\
\midrule Energy & 10 & 0 & 0 & 0  \\
Materials & 6 & 0 & 4 & 0  \\
Industrials & 5 & 2 & 0 & 3  \\
Automobiles and Auto Parts & 10 & 0 & 0 & 0 \\
Durables and Apparel & 4 & 1 & 0 & 5  \\
Consumer Service & 7 & 0 & 3 & 0 \\
Media & 1 & 1 & 8 & 0  \\
Retail & 1 & 3 & 5 & 0  \\
Consumer Staples & 1 & 8 & 1 & 0  \\
Healthcare & 0 & 10 & 0 & 0  \\
Financials & 5 & 0 & 5 & 0  \\
Information Technology & 4 & 0 & 2 & 4  \\
Utilities & 0 & 10 & 0 & 0 \\
Real Estate & 1 & 7 & 1 & 1  \\
\bottomrule

   \end{tabular}
\label{Group structure}
\end{table}

\begin{table}[!htbp]
    \centering
    \caption{Estimation of column loading matrix. The loading matrix are multiplied by 30 and rounded to integers for ease in display.}
    \begin{tabular}{lcccccccccc}
    \toprule
    \textbf{Column Factor} & CNM & HK & US & UK & JP  \\
    \midrule
    Factor 1 & 0 & 8 & -2 & 13 & 26  \\
    \bottomrule
    \end{tabular}
    \label{Column loading matrix}
\end{table}

\begin{table}[!htbp]
    \centering
    \caption{Estimation of column loading matrix (matrix factor model). The loading matrix are multiplied by 30 and rounded to integers for ease in display.}
    \begin{tabular}{lcccccccccc}
    \toprule
    \textbf{Column Factor} & CNM & HK & US & UK & JP  \\
    \midrule
    Factor 1 & 2 & 4 & 25 & 17 & 2  \\
    \bottomrule
    \end{tabular}
    \label{Column loading matrix2}
\end{table}

Finally, we define out of sample $R^2$ on a testing set of size $T$ as
\begin{equation}
R^2  \triangleq 1-\frac{\sum_{t=1}^T\left\|\mathbf{Y}_t-\widehat{\mathbf{Y}}_t\right\|_F^2}{\sum_{t=1}^T\left\|\mathbf{Y}_t-\overline{\mathbf{Y}}\right\|_F^2},
\end{equation}
where $\overline{\mathbf{Y}}=\frac{1}{T} \sum_{t=1}^T \mathbf{Y}_t$ and $\widehat{\mathbf{Y}}_t$ represents the fitted values computed through the model. Table \ref{Model Evaluation Results} presents a direct comparison between our model and matrix factor models with varying numbers of factors. We use $R^2_{K}$, $R^2_{U}$ and  $R^2_{T}$ to represent the out of sample $R^2$ of known factors term, unknown factors term and the total of these two terms in our model, respectively. The known factors term accounts for 24.5\% of the total variation in our model. Even we use the high dimensions of unknown factors, the variances explained by the known factors term in our model are larger than that explained by the unknown factors term in the matrix factor model. This finding proves that it is very effective to introduce known factors term $\mathbf{A} \mathbf{X}_t$ based on matrix factor model. 
%
\begin{table}[]
\centering
\caption{Model Evaluation Results.}
\begin{tabular}{c|ccc|c}
\hline & \multicolumn{3}{|c|}{ Matrix regression with latent factors } & Matrix Factor Model \\
\hline Factors & $R^2_{K}$ & $R^2_{U}$ & $R^2_{T}$ & $R^2$ \\
\hline$(2,1)$ & & 0.036 & 0.289 & 0.143 \\
$(5,1)$ & & 0.045 & 0.300 & 0.159 \\
$(10,1)$ & 0.245 & 0.056 & 0.310 & 0.173 \\
$(20,1)$ & & 0.072 & 0.326 & 0.190 \\
$(30,1)$ & & 0.084 & 0.338 & 0.202 \\
$(40,1)$ & & 0.095 & 0.350 & 0.212 \\
\hline
\end{tabular}
\label{Model Evaluation Results}
\end{table}

\section{Conclusion}
\noindent{}In this article, we incorporate the known factors term into the traditional matrix factor model. The known factors term can be estimated by the least square methods. Subsequently, we apply the estimation procedure in \citet{wang2019factor} to estimate the unknown factor loadings and the dimensions of unknown factors. The convergence rates we establish align with those specified in \citet{wang2019factor}. The numerical studies show the performance of our estimation methods in finite samples. Furthermore, the real data analysis underscores the necessity of introducing the known factor term.

\bibliographystyle{cas-model2-names}

\bibliography{arxivversion}

\end{document}